\documentclass[useAMS,usedcolumn,usegraphicx,usenatbib]{mn2e}

\usepackage{times}
\usepackage{amsmath}                  
\usepackage{amssymb}                  
\usepackage{graphicx}
\usepackage{units}
\usepackage{dcolumn}
\usepackage{longtable}
\usepackage{lscape}
\usepackage{booktabs}
\usepackage{natbib}
\usepackage{aas_macros}
\usepackage[draft,pdftex,pdfpagemode={UseOutlines},bookmarks,bookmarksopen,colorlinks,linkcolor={blue},citecolor={green},urlcolor={red}]{hyperref}

\makeatletter
\newcolumntype{d}[1]{>{\DC@{,}{{.}}{#1}}c<{\DC@end}}
\newcolumntype{o}[1]{>{\DC@{+}{\pm}{#1}}c<{\DC@end}}
\makeatother
\makeatletter
\newcolumntype{f}[1]{>{\DC@{p}{\ldots}{#1}}c<{\DC@end}}
\makeatother

\let\orgautoref\autoref
\providecommand{\Autoref}
        {\def\equationautorefname{Equation}%
         \def\figureautorefname{Fig.}%
         \def\subfigureautorefname{Fig.}%
         \def\partautorefname{Part}%
         \def\chapterautorefname{Chapter}%
         \def\sectionautorefname{Section}%
         \def\subsectionautorefname{Section}%
         \def\subsubsectionautorefname{Section}%
         \def\Itemautorefname{Item}%
         \def\tableautorefname{Table}%
         \def\lstlistingautorefname{Listing}%
         \orgautoref}

\renewcommand{\autoref}
        {\def\equationautorefname{equation}%
         \def\figureautorefname{Fig.}%
         \def\subfigureautorefname{Fig.}%
         \def\partautorefname{part}%
         \def\chapterautorefname{chapter}%
         \def\sectionautorefname{section}%
         \def\subsectionautorefname{section}%
         \def\subsubsectionautorefname{section}%
         \def\appendixautorefname{appendix}%
         \def\Itemautorefname{item}%
         \def\tableautorefname{Table}%
         \def\lstlistingautorefname{Listing}%
         \orgautoref}
    
\makeatletter

\newcommand{\rxjn}{RX\,J0720.4-3125}
\newcommand{\psrjon}{PSR\,J0953$+$0755}
\newcommand{\psrjns}{PSR\,J0630$-$2834}
\newcommand{\dsunheute}{d_{\odot,today}}
\newcommand{\dsunSN}{d_{\odot,SN}}
\newcommand{\commacirc}{^\circ\hspace{-0.8ex}.}

\title[The Neutron Star Born in the Antlia SNR]{The Neutron Star Born in the Antlia Supernova Remnant}
\author[N. Tetzlaff et al.]{N. Tetzlaff$^{1}$\thanks{E-mail:
nina@astro.uni-jena.de}, G. Torres$^{2}$, R. Neuh\"auser$^{1}$ and M. M. Hohle$^{1}$\\
$^{1}$Astrophysikalisches Institut und Universit\"ats-Sternwarte Jena, Schillerg\"asschen 2-3, 07745 Jena, Germany\\
$^{2}$Harvard-Smithsonian Center for Astrophysics, 60 Garden St., Cambridge, MA 02138, USA}

\begin{document}

\date{Accepted. Received; in original form}

\pagerange{\pageref{firstpage}--\pageref{lastpage}} \pubyear{2013}

\maketitle

\label{firstpage}

\begin{abstract} 
Among all known young nearby neutron stars, we search for the neutron star that was born in the same supernova event that formed the Antlia supernova remnant (SNR). We also look for a runaway star that could have been the former companion to the neutron star (if it exists) and then got ejected due to the same supernova.

We find the pulsar \psrjns{} to be the best candidate for a common origin with the Antlia SNR. In that scenario the SNR is $\approx\unit[1{.}2]{Myr}$ old and is presently located at a distance of $\approx\unit[138]{pc}$. We consider the runaway star HIP 47155 a former companion candidate to \psrjns{}. The encounter time and place is consistent with both stars being ejected from the Antlia SNR. We measured the radial velocity of HIP 47155 as $\unit[32{.}42\pm0{.}70]{km/s}$.
\end{abstract}



\section{Introduction}

The Antlia supernova remnant (SNR), located at $\left(l,b\right)=\left(276\commacirc5,19^\circ\right)$, was first discovered by \citet{2002ApJ...576L..41M} in X-ray observations. It was later confirmed as SNR in the ultraviolet \citep{2007ApJ...670.1132S}. Its observability in the UV range as well as its large projected diameter of $\approx\unit[24]{^\circ}$ suggests small distance (up to a few hundred pc). \citet{2002ApJ...576L..41M} estimated the distance to the Antlia SNR as $d_{A}\lesssim\unit[500]{pc}$ with a preference for smaller distances ($\approx\unit[100]{pc}$). They assessed an SNR age of $\approx\unit[2]{Myr}$ supported by their suggestion that the pulsar \psrjon{} and the Antlia SNR share a common origin. Given the detection of $^{60}$Fe in the Earth's crust of nearly $\approx\unit[2]{Myr}$ old \citep{2004PhRvL..93q1103K,2008PhRvL.101l1101F}, \citet{2002ApJ...576L..41M} also considered whether this could have been formed in that supernova. However, \citet{2002ApJ...576L..41M} only considered eight nearby pulsars listed by \citet{2001A&A...365...49H}. Therefore, it is worthwhile to re-visit this issue. 

Due to large uncertainties in the distances of neutron stars (NSs) and their (in most cases) unknown radial velocities, multiple possible birth places (young associations and clusters, isolated SNRs) can usually be associated \citep{2010MNRAS.402.2369T,2011MNRAS.417..617T}. Therefore, further indicators are needed to decide on a particular birth place. A promising indicator is the identification of a possible former companion that is now a so-called runaway star \citep{1961BAN....15..265B}. These are typically fast-moving young single stars that may show signs of former binary evolution such as high helium abundance and high rotational velocity due to mass and momentum transfer from the primary as it filled its Roche lobe as well as possibly enhanced $\alpha$ elements as supernova debris material. 

By constructing the past flight paths of young NSs and young runaway stars, we aim on identifying the NS that was born in the Antlia SNR. For those NSs for which we consider an association with the Antlia SNR possible, we also investigate whether a different origin in a nearby young association or cluster is possible. To account for the errors on the observables as well as the for NSs unknown radial velocity, we perform Monte Carlo simulations. 

With this work we aim to extend the analysis done by \citet{2002ApJ...576L..41M} by investigating a larger sample of neutron stars and, as an additional indicator, search for a former companion candidate to the neutron star that was then ejected in the same supernova event that formed the Antlia SNR.

We describe our method in \autoref{sec:Method} before presenting the analysis and results in \autoref{sec:results} and giving concluding remarks in \autoref{sec:concl}.


\section{Method}\label{sec:Method}

We use the same approach as already described in \citet{2010MNRAS.402.2369T} and \citet{2011MNRAS.417..617T} (see also \citealt{2001A&A...365...49H}). Therefore, only a brief description is given here.

To construct the trajectories of the Antlia SNR, young NSs and runaway stars (and young associations and clusters, \citealt{2010MNRAS.402.2369T,2012PASA...29...98T,2013PhD...Nina}), we apply Monte Carlo simulations by varying the observables (parallax, proper motion, radial velocity) within their error intervals. For the radial velocity of the NS, we assume a probability distribution such that the distribution of pulsar space velocities according to \citet{2005MNRAS.360..974H} is satisfied. Runaway star data were taken from the \citet{2011MNRAS.410..190T} catalogue for Hipparcos runaway stars (also \citealt{2013PhD...Nina}). If the radial velocity of a runaway star is unknown, we vary it randomly within $\pm\unit[500]{km/s}$.

In a time range between $0$ (today) and $\unit[5\times10^6]{yr}$ (into the past, in steps of $\unit[10^4]{yr}$) the past separation between the centre of the Antlia SNR (adopting $\left(l,b\right)=\left(276\commacirc5,19^\circ\right)$, \citealt{2002ApJ...576L..41M}) and the NS is evaluated. Later, also runaway stars are traced back simultaneously. The smallest separation $d_{min}$ for each pair of trajectories (Antlia SNR and NS or runaway star and NS) and the associated time $\tau$ in the past is stored. The distribution of separations $d_{min}$ is supposed to obey a distribution of absolute differences of two 3D Gaussians \citep[see][]{2001A&A...365...49H,2010MNRAS.402.2369T,2011MNRAS.417..617T}.
%
%

%
%

The distance to the Antlia SNR is uncertain. For the Monte Carlo simulations, we adopt values of $\unit[100^{+400}_{-40}]{pc}$ to take into account the lower and upper limits by \citet{2002ApJ...576L..41M} and that they prefer a distance of $\approx\unit[100]{pc}$.

The actual (observed) case is different from this simple model (no 3D Gaussian distributed positions, due to e.g. the Gaussian distributed parallax that goes into the position reciprocally, complicated radial velocity distribution, etc.). Therefore, we will adapt the theoretical formulae (equations 1 and 2 in \citealt{2012PASA...29...98T}; here we use the symbols $\mu$ and $\sigma$ for the expectation value and standard deviation, respectively) only to the first part of the $d_{min}$ distribution (up to the peak plus a few more bins, see \citealt{2012PASA...29...98T}). The derived parameter $\mu$ then gives the positional difference between the two objects. Typically, a few million trajectories are constructed throughout a Monte Carlo simulation.\\

This procedure was already successfully applied by \citet{2001A&A...365...49H}, \citet{2008AstL...34..686B,2009AstL...35..396B} and us \citep{2009MNRAS.400L..99T,2010MNRAS.402.2369T,2011MNRAS.417..617T,2012PASA...29...98T}. An investigation of (artificial) test cases showed that it is well possible to recover place and time of the formation of a NS \citep{2013PhD...Nina}.


\section{Results}\label{sec:results}

\subsection{Search for the related neutron star}

\begin{figure}
\centering
\includegraphics[width = 0.45\textwidth,viewport = 40 205 540 600]{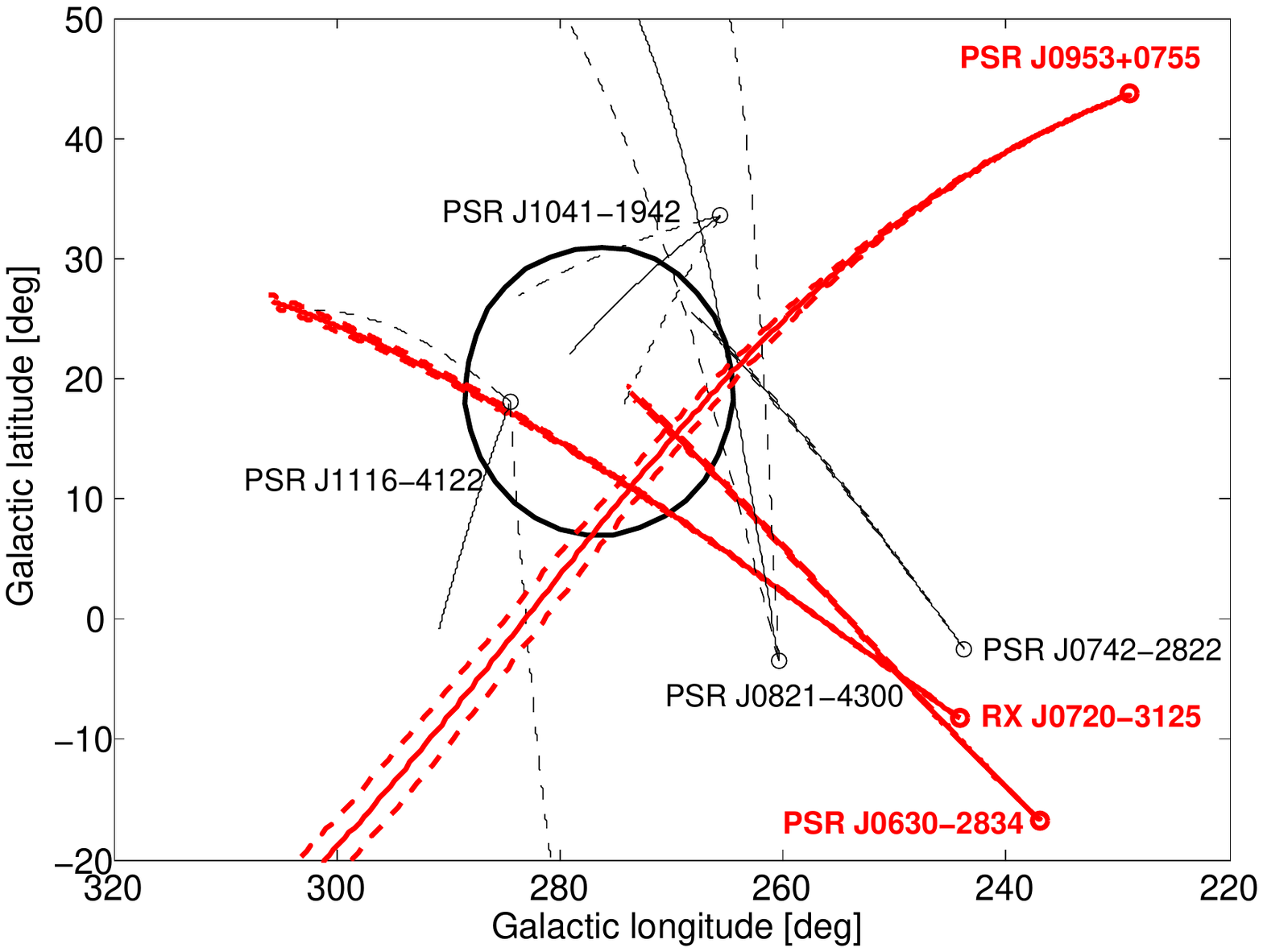}
\caption[The Antlia SNR and past paths of seven NSs.]{The Antlia SNR (thick circle) and the projected past flight paths of seven NSs (calculated back in time for $\unit[5]{Myr}$; flight paths shown here for $v_r=\unit[0]{km/s}$ for six stars and $v_r=\unit[100]{km/s}$ for \psrjon{}, as \citet{2002ApJ...576L..41M}; dashed lines indicate the $1\sigma$ error on the proper motion). The four NSs drawn in black are too distant ($\approx\unit[2]{kpc}$) for a common origin with the Antlia SNR. The three NSs drawn in red (thick lines) are candidates for a common origin with the Antlia SNR.}\label{fig:NScrossAntlia}
\end{figure}

Among all young (spin-down ages smaller than $\unit[50]{Myr}$) nearby ($\lesssim\unit[3]{kpc}$) NSs with known proper motion, 106 in total (ATNF pulsar database\footnote{http://www.atnf.csiro.au/research/pulsar/psrcat/}, \citealt{2005AJ....129.1993M}), the projected past paths of seven NSs cross the Antlia SNR during the past $\unit[5]{Myr}$, \autoref{fig:NScrossAntlia}. 
Four of them are too distant ($\approx\unit[2]{kpc}$) and were not closer than a few hundred pc to the remnant for any reasonable radial velocity. Among the other three is \rxjn{} which was probably born in the Trumpler 10 association $\approx\unit[1]{Myr}$ ago \citep{2003AA...408..323M,2007ApJ...660.1428K,2010MNRAS.402.2369T,2011MNRAS.417..617T} but still should be considered as a candidate for an origin in the Solar neighbourhood or the Antlia SNR (as it could be nearby, \citealt{2002ApJ...576L..41M}). The remaining two are \psrjns{} and \psrjon{}. The properties of \psrjns{}, \rxjn{} and \psrjon{} are given in \autoref{tab:properties}. 
\begin{table*}
\centering
\caption{Parameters of \psrjns{}, \rxjn{} and \psrjon{}. \newline
$\alpha$, $\delta$: equatorial coordinates (J2000.0), $\pi$: parallax (radio parallaxes for \psrjns{} and \psrjon{}, optical parallax for \rxjn{}), $\mu^*_\alpha=\mu_\alpha\cos\delta$, $\mu_\delta$: proper motion in right ascension and declination, respectively.}\label{tab:properties}
\begin{tabular}{l c c c c c l}
\toprule
 & $\alpha$ (h:m:s) & $\delta$ (d:m:s) & $\pi$ (mas) & $\mu^*_{\alpha}$ (mas yr$^{-1}$) & $\mu_\delta$ (mas yr$^{-1}$) & Refs. \\\midrule
\psrjns{} & 06:30:49.4043 & -28:34:42.78 & $3{.}01\pm0{.}41$ & $-46{.}30\pm0{.}99$ & $21{.}26\pm0{.}52$ & 1 \\
\rxjn{} &   07:20:24.9620 & -31:25:50.08 & $3{.}6\pm1{.}6$   & $-92{.}8\pm1{.}4$   & $55{.}3\pm1{.}7$   & 2, 3\\
\psrjon{} & 09:53:09.3097 & +07:55:35.75 & $3{.}82\pm0{.}07$ & $-2{.}09\pm0{.}08$  & $29{.}46\pm0{.}07$ & 4, 5\\
\bottomrule
\end{tabular}
\newline
References: 1 -- \citet{2009ApJ...701.1243D}, 2 -- \citet{2007ApJ...660.1428K}, 3 -- \citet{ThomasPhD}, 4 -- \citet{2004MNRAS.353.1311H}, 5 -- \citet{2002ApJ...571..906B}
\end{table*}

To check whether nearby associations or clusters could host the birth places of \psrjns{} and \psrjon{}\footnote{For \rxjn{} this analysis was already carried out by \citealt{2010MNRAS.402.2369T,2011MNRAS.417..617T}. Thereafter, \rxjn{} was probably born in the Trumpler 10 association.}, we investigated close encounters with young nearby associations and clusters \citep{2010MNRAS.402.2369T,2012PASA...29...98T,2013PhD...Nina}. The past trajectories of both stars point to several nearby young local associations \citep{2008A&A...480..735F,2008hsf2.book..757T}; however, no convincing birth association could be found. We consider an origin in the Solar neighbourhood or the Antlia SNR most likely for these stars.\\

\begin{figure}
\centering
\hspace*{-2em}\includegraphics[width=0.55\textwidth]{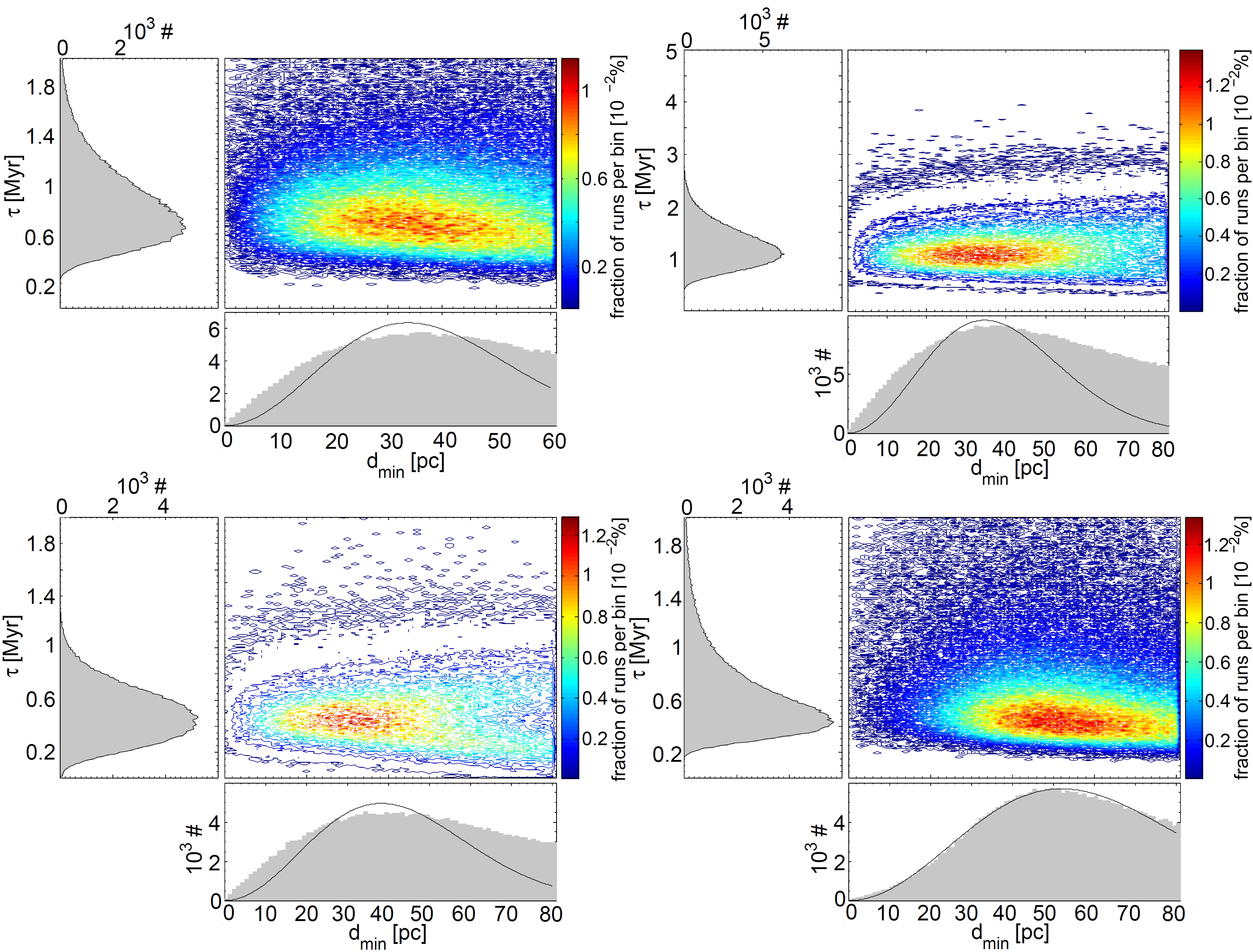}
\caption[$d_{min}$ and $\tau_{kin}$ distributions for the Antlia SNR and three NSs.]{Distributions of minimum separations $d_{min}$ and corresponding flight times $\tau$ for encounters between the Antlia SNR and three NSs, \psrjns{} (top panel), \rxjn{} (bottom left panel) and \psrjon{} (bottom right panel). For \psrjns{} the distributions using a radial velocity distribution derived from pulsar spatial velocities \citep{2005MNRAS.360..974H} (top left panel) as well as a Gaussian distribution with $v_r = \unit[200\pm100]{km/s}$ (top right) are shown (see text). The solid curves drawn in the $d_{min}$ histograms (bottom panels) represent the theoretically expected distributions (equation 2 in \citealt{2012PASA...29...98T}), adapted to the first part of each histogram: For \psrjns{} $\mu=0$, $\sigma=\unit[23{.}9]{pc}$ (top left) and $\sigma = \unit[24{.}5]{pc}$ (top right); for \rxjn{} $\mu=0$, $\sigma=\unit[27{.}1]{pc}$ (bottom left); for \psrjon{} $\mu=0$, $\sigma=\unit[36{.}9]{pc}$ (bottom right). The encounter times are $\tau=\unit[0{.}64^{+0{.}36}_{-0{.}21}]{Myr}$ (top left) and $\tau=\unit[1{.}08^{+0{.}40}_{-0{.}31}]{Myr}$ (top right) for \psrjns{}, $\tau=\unit[0{.}47^{+0{.}20}_{-0{.}22}]{Myr}$ for \rxjn{} (bottom left) and $\tau=\unit[0{.}43^{+0{.}31}_{-0{.}15}]{Myr}$ for \psrjon{} (bottom right), respectively. One of these stars might be associated with the Antlia SNR.}\label{fig:contourplushistsAntlia}
\end{figure}

The past separation between the Antlia SNR and the three NSs \psrjns{}, \psrjon{} and \rxjn{} were then evaluated, \autoref{fig:contourplushistsAntlia}.\footnote{The Antlia remnant was assumed to be moving on a constant orbit around the Galactic centre.} 

For \psrjon{} equation 2 in \citealt{2012PASA...29...98T} ($\mu=0$) fits well the $d_{min}$ distribution. Note that the obtained encounter time of $\approx\unit[0{.}5]{Myr}$ is considerably smaller than the one claimed by \citet{2002ApJ...576L..41M} ($\tau\approx\unit[2-4]{Myr}$). An encounter time of $\approx\unit[2]{Myr}$ is obtained if $v_r=\unit[50\pm50]{km/s}$ is assumed for the pulsar (comparable to the range \citealt{2002ApJ...576L..41M} adopted). However, then a theoretical curve with $\mu=\unit[48{.}2]{pc}$ and $\sigma=\unit[25{.}8]{pc}$ fits well the $d_{min}$ distribution rather than $\mu=0$. It is still possible that \psrjon{} was inside the Antlia SNR, though less likely. Moreover, the resulting space velocity of the pulsar would be very small, $\approx\unit[80]{km/s}$, which is unlikely but not impossible. For the calculation using a reasonable space velocity distribution for \psrjon{} \cite[from][]{2005MNRAS.360..974H}, the present NS parameters and position of the encounter are given in \autoref{tab:predparAntlia}. As already expected from \autoref{fig:NScrossAntlia}, the predicted encounter position of \psrjon{} is only marginally consistent with the observed position of the Antlia SNR.\\
\begin{table*}
\centering
\caption[NSs possibly associated with the Antlia SNR.]{Present-day parameters and encounter position and time for three NSs that might be associated with the Antlia SNR. \newline
Column 2: predicted separation between the encounter and the centre of the Antlia SNR (expectation value $\mu$ and standard deviation $\sigma$ inferred from equations 1 or 2 in \citealt{2012PASA...29...98T}). Column 3: encounter time $\tau$. Columns 4-8: Predicted present NS parameters (heliocentric radial velocity $v_r$, proper motion $\mu_\alpha^*$ and $\mu_\delta$, peculiar space velocity $v_{sp}$, parallax $\pi$). Note that it is possible that the derived value for $v_r$ is larger than that of $v_{sp}$ because $v_r$ is heliocentric whereas $v_{sp}$ is the peculiar velocity of the NS that reflects its kick velocity. Columns 9-12: Predicted supernova position (supernova distance $\dsunSN$, at the time of the supernova; present distance $\dsunheute$ and Galactic coordinates, $l$ and $b$, J2000.0 of the centre of the SNR). Error bars denote $\unit[68]{\%}$ confidence (for the derivation of the parameters we refer to \citealt{2010MNRAS.402.2369T}). For \psrjns{} the results using a radial velocity distribution derived from the pulsar spatial velocities \citep{2005MNRAS.360..974H} ($^*$) as well as a Gaussian distribution with $v_r=\unit[200\pm100]{km/s}$ ($^\#$) are shown (see text).}\label{tab:predparAntlia}
\setlength\extrarowheight{4pt}
\begin{tabular}{@{}c<{\hspace{1ex}}@{} >{$}c<{$\hspace{1.5ex}}@{} >{$}r<{$\hspace{1.5ex}}@{} | >{$\ }r<{$\hspace{1ex}}@{} o{4.2}<{\hspace{1ex}}@{} o{4.2}<{\hspace{1ex}}@{} >{$}r<{$\hspace{1ex}}@{} >{$}r<{$\hspace{1.5ex}}@{} | >{$}c<{$\hspace{1ex}}@{} >{$}c<{$\hspace{1ex}}@{} >{$}c<{$\hspace{1ex}}@{} >{$}c<{$}@{}}
\toprule
NS	&	\multicolumn{1}{c}{$\left(\mu,\sigma\right)$} & \multicolumn{1}{c|}{$\tau$} &	\multicolumn{5}{c|}{Predicted present-day NS parameters}	&	\multicolumn{4}{c}{Predicted supernova/SNR position} \\ 
	&\multicolumn{1}{c}{}	&  &	\multicolumn{1}{c}{$v_r$}			& \multicolumn{1}{c}{$\mu_{\alpha}^*$} & \multicolumn{1}{c}{$\mu_{\delta}$} & \multicolumn{1}{c}{$v_{sp}$} &	\multicolumn{1}{c|}{$\pi$}			&	\dsunSN & \dsunheute 	& \multicolumn{1}{c}{$l$}	& \multicolumn{1}{c}{$b$}  \\ 
		& \multicolumn{1}{c}{[pc]} & \multicolumn{1}{c|}{[Myr]}	& \multicolumn{1}{c}{[km/s]} & \multicolumn{1}{c}{[mas/yr]} & \multicolumn{1}{c}{[mas/yr]} & \multicolumn{1}{c}{[km/s]} & \multicolumn{1}{c|}{[mas]} & \multicolumn{1}{c}{[pc]} & \multicolumn{1}{c}{[pc]} & \multicolumn{1}{c}{[$^\circ$]} & \multicolumn{1}{c}{[$^\circ$]} \\\midrule
\psrjns{}$^*$					& \left(0,23{.}9\right)	& 0{.}64^{+0{.}36}_{-0{.}21}	& 365^{+136}_{-94} & -46{.}3+1{.}0 & 21{.}2+0{.}5 & 375^{+131}_{-84} & 2{.}9^{+0{.}5}_{-0{.}3}	& 62^{+24}_{-17}	&	72^{+21}_{-26}	& 268{.}4^{+17{.}6}_{-14{.}0}	& 16{.}7^{+14{.}2}_{-12{.}7} \\
\psrjns{}$^\#$					& \left(0,24{.}5\right)	& 1{.}08^{+0{.}40}_{-0{.}31}	& 212^{+103}_{-54} & -46{.}3+1{.}0 & 21{.}3+0{.}5 & 235^{+61}_{-87} & 3{.}1^{+0{.}4}_{-0{.}4}	& 102^{+35}_{-33}	&	105^{+40}_{-36}	& 269{.}7^{+9{.}4}_{-7{.}8}	& 18{.}6^{+7{.}4}_{-7{.}7} \\
\rxjn{}			& \left(0,27{.}1\right)& 0{.}47^{+0{.}20}_{-0{.}22}	& 432^{+92}_{-187}& -92{.}8+1{.}4	& 55{.}3+1{.}7	& 352^{+172}_{-100} & 4{.}1^{+0{.}8}_{-1{.}4}& 67^{+56}_{-22}	&	75^{+47}_{-35}	& 269{.}9^{+16{.}4}_{-10{.}2}	& 10{.}9^{+9{.}9}_{-6{.}2} \\
\psrjon{}		& \left(0,36{.}9\right)	& 0{.}43^{+0{.}31}_{-0{.}15}	& 442^{+88}_{-139}	& -2{.}1+0{.}1	& 29{.}5+0{.}1	& 435^{+86}_{-139} & 3{.}8^{+0{.}1}_{-0{.}1}& 50^{+28}_{-17} &	57^{+17}_{-28} & 250-290^\mathsf{a}	& -30$ to $+20^\mathsf{a} \\
\bottomrule
\multicolumn{12}{p{0.9\textwidth}}{$^\mathsf{a}$ For \psrjon{} the distributions for $l$ and $b$ are very broad with no clear peak, thus an interval is given.}
\end{tabular}
\end{table*}

Although the $d_{min}$ distributions in the cases \psrjns{} and \rxjn{} are not well represented by equations 1 and 2 in \citealt{2012PASA...29...98T} (probably because the parallax error is large in both cases, $\unit[14]{\%}$ for \psrjns{}, $\unit[44]{\%}$ for \rxjn{}, whereas for \psrjon{} it is only $\unit[2]{\%}$), they suggest that both objects could have been at the same place (as the supernova) at the same time in the past. In the case of \rxjn{}, the predicted encounter position is again only marginally consistent with the observed SNR centre (also seen in \autoref{fig:NScrossAntlia}). Together with the previous result that \rxjn{} was probably born in the Trumpler 10 association, we consider a common origin of this NS with the Antlia SNR less likely. 

For \psrjns{}, the encounter position is the most consistent with the observed coordinates of the Antlia SNR centre, making it the best candidate for the pulsar that can be associated with the Antlia SNR. The inferred ages of the pulsar and the SNR if they originated from the same supernova event are $\approx\unit[0{.}6]{Myr}$.

The small supernova distance of $\approx\unit[45-90]{pc}$ implies an absolute SNR radius of $\approx\unit[10-20]{pc}$. For an $\unit[0{.}4-1{.}0]{Myr}$ old SNR, SNR expansion theory (Snowplough expansion, e.g. \citealt{1982SvAL....8..361B}) predicts a radius of $\approx\unit[45-55]{pc}$ for an ISM density of $n=\unit[1]{cm^{-3}}$ and standard explosion energy of $E=\unit[10^{51}]{erg}$. Either the explosion energy was considerably smaller ($\approx\unit[10^{49}]{erg}$) or the supernova occurred at a somewhat larger distance (also implying a slightly larger age). Theoretical core-collapse supernova models do not predict explosion energies $<\unit[4\cdot10^{50}]{erg}$ \citep{2012ApJ...757...69U}. For a larger supernova distance ($\approx\unit[100-200]{pc}$ such that the predicted size of the SNR is consistent with the observed one at an age of $\approx\unit[1-2]{Myr}$), the radial velocity of \psrjns{} is required to be smaller ($\approx\unit[100-300]{km/s}$), inferring a spatial NS velocity of $\approx\unit[100-300]{km/s}$ (kick velocity). 

Assuming NS radial velocities of $\unit[200\pm100]{km/s}$ yields a distribution $d_{min}$ that is still consistent with equation 2 in \citealt{2012PASA...29...98T} ($\mu=0$, $\sigma=\unit[24{.}5]{pc}$). The time of the encounter is $\approx\unit[0{.}7-1{.}4]{Myr}$ in the past. The predicted distance to the supernova is $\approx\unit[70-150]{pc}$ (as seen from Earth today). The corresponding absolute SNR radius is $\approx\unit[15-30]{pc}$. This is in marginal agreement with the theoretically expected value of $\approx\unit[30-60]{pc}$. \\

\subsection{The former companion candidate to \psrjns{}}

The scenario in which the pulsar \psrjns{} is the compact remnant of the Antlia SNR is supported by the identification of the former companion candidate HIP 47155 for which the encounter position with \psrjns{} coincides with the Antlia SNR. The radial velocity of HIP 47155 (spectral type A5/7(IIw), \citealt{1982mcts.book.....H}; kA3hF2mA3V, \citealt{2001A&A...373..625P}) was previously unknown. Its measurement is crucial to confirm or reject HIP 47155 as former companion candidate to \psrjns{}. HIP 47155 was considered a runaway star by \citet{2011MNRAS.410..190T} due to its high transverse velocity of $\unit[31^{+6}_{-2}]{km/s}$.\\

\begin{figure}
\centering
\includegraphics[width = 0.45\textwidth]{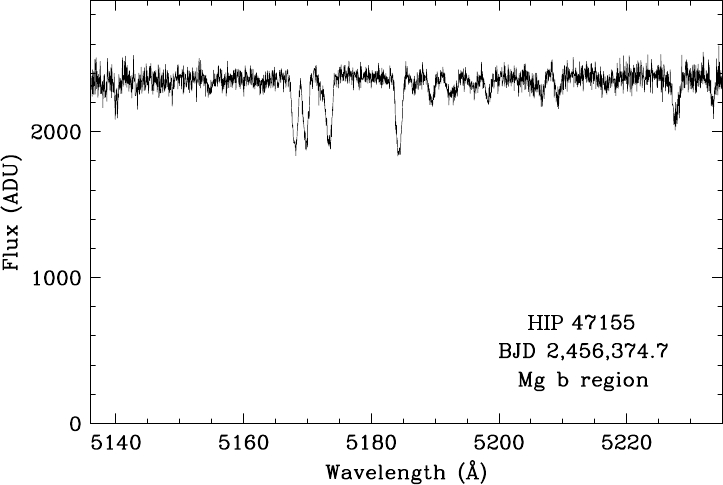}
\caption{Spectrum of HIP\,47155 on BJD 2,456,374.7068,
showing echelle order \#23 centered on the lines of the Mg I\,b
triplet. This order carries most of the velocity information.}\label{fig:specHIP47155}
\end{figure}

Spectroscopic observations of HIP\,47155 were gathered using the TRES
instrument \citep{Furesz08} on the 1.5\,m Tillinghast reflector at
the F.\ L.\ Whipple Observatory (Mount Hopkins, Arizona, USA). Three
exposures were obtained over a period of 25 days in April and May of
2013. Radial velocity standard stars were observed each night with the same setup to monitor the velocity zero point of the instrument. The spectra cover the wavelength range $\sim$3900--8900\,\AA,
and were taken at a typical resolving power of $R \approx 44,\!000$
with the medium fiber of the instrument. The signal-to-noise ratios
for the individual 3- to 5-minute exposures range between 80 and 100
per resolution element of $\unit[6{.}8]{km/s}$, and refer to the region of the
Mg I\,b triplet ($\sim$5200\,\AA). All spectra were reduced
and extracted using standard procedures as described by \citet{Buchhave:10a} and \citet{Buchhave:10b}. These templates cover a $\unit[300]{\AA}$ window centred at $\unit[5200]{\AA}$, which generally contains the most information on the velocity of the star. \Autoref{fig:specHIP47155} shows a portion of one of our observations.

Radial velocities were derived by cross-correlation against a
synthetic template taken from a large library computed by John Laird,
based on model atmospheres by R.\ L.\ Kurucz and a line list developed
by Jon Morse \citep[see][]{2012Natur.486..375B}. The task of selecting the
optimum template was somewhat complicated by the peculiar chemical
composition of the star, which is considered to be a member of the
$\lambda$\,Boo class with a spectral type of kA3hF2mA3V \citep{2001A&A...373..625P}. To establish the best template we adopted the metallicity
and surface gravity derived by \citet{2011A&A...530A.138C}, which are
${\rm [Fe/H]} = -1.13$ and $\log g = 3.67$, and explored a range of
temperatures ($T_{\rm eff}$) and rotational broadenings ($v \sin i$
when seen in projection), which are the two parameters that affect the
velocities the most. The best match to the observed spectra was found
for $T_{\rm eff} = \unit[8250]{K}$ and $v \sin i = \unit[45]{km/s}$. The temperature
is consistent with the above $\lambda$\,Boo classification for the
metallic lines (A3). The weighted average of the three velocity
measurements is $\unit[+32{.}42\pm0{.}70]{km/s}$, and the individual values (in
the heliocentric frame) with corresponding uncertainties are listed in
\autoref{tab:specHIP47155}. The above average velocity is robust against changes in the template parameters. For example, changing the temperature within one step in our grid ($\unit[250]{K}$) leads to velocity changes that are well within the quoted uncertainties.\\
\begin{table}
\centering
\caption{Heliocentric radial velocities RV and their $1\sigma$ errors err for HIP 47155, both in km/s.}\label{tab:specHIP47155}
\begin{tabular}{d{4} d{2} d{2}}
\toprule
\multicolumn{1}{c}{BJD}    &  \multicolumn{1}{c}{RV} & \multicolumn{1}{c}{err}\\\midrule
	2456374.7068  & +31.51 &  1.29 \\
  2456386.7075  & +32.83 &  1.19 \\
  2456399.6398  & +32.75 &  1.15 \\\bottomrule
\end{tabular}
\end{table}

\begin{figure}
\centering
\includegraphics[width=0.45\textwidth, viewport= 20 210 580 625]{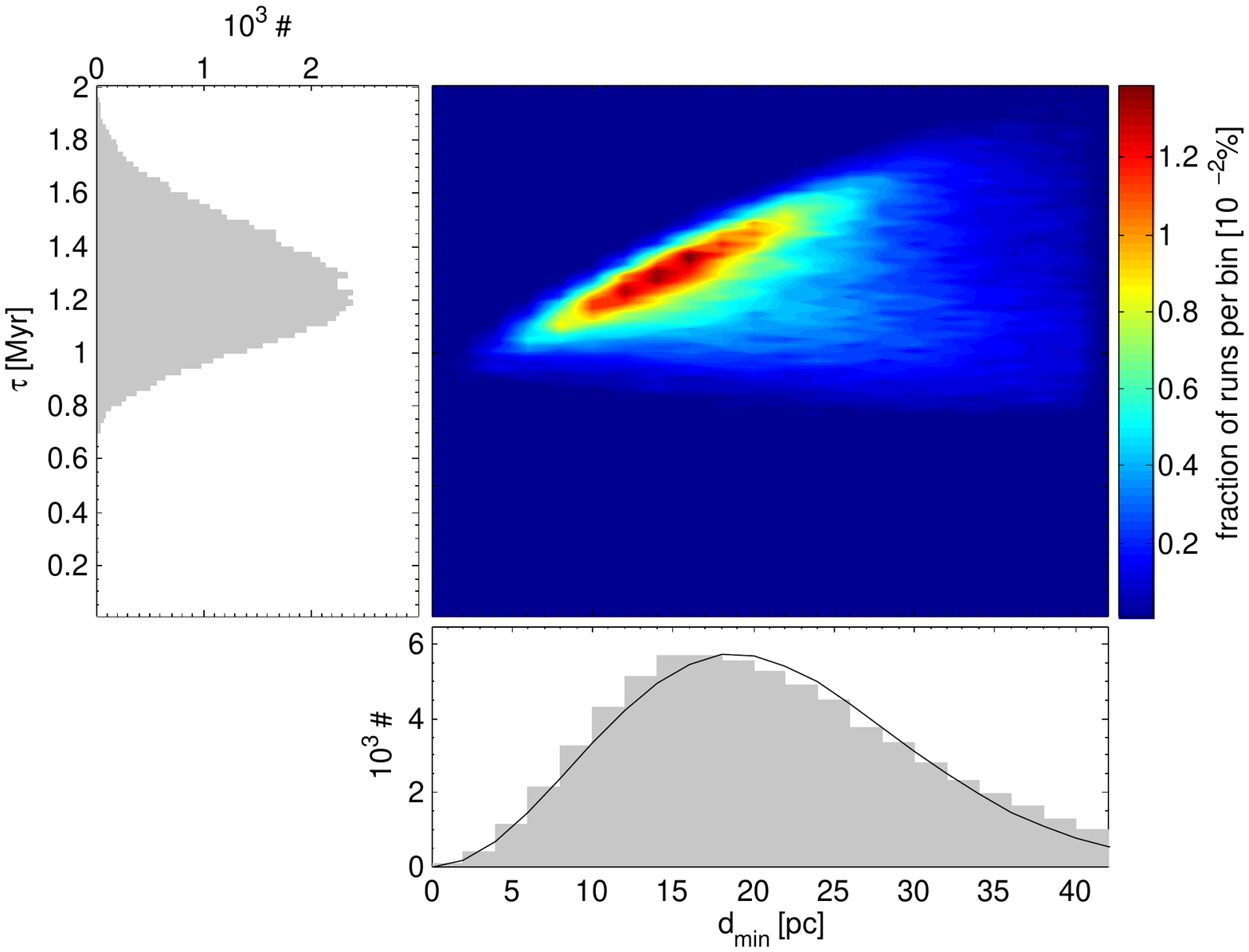}
\caption{Distribution of minimum separations $d_{min}$ and corresponding flight time $\tau$ for encounters between \psrjns{} and HIP 47155 for those runs where both stars were inside the Antlia SNR (assuming spherical shape with a radius of $\unit[{30}]{pc}$). For \psrjns{} a Gaussian $v_r$ distribution with $v_r=\unit[200\pm100]{km/s}$ is adopted. The solid curves drawn in the $d_{min}$ histograms (bottom panel) represent the theoretically expected distribution (equation 2 in \citealt{2012PASA...29...98T}): $\mu=0$, $\sigma=\unit[{13{.}3}]{pc}$. The flight time of the runaway star since ejection is $\tau=\unit[{1{.}20^{+0{.}26}_{-0{.}18}}]{Myr}$.}\label{fig:0630HIP}
\end{figure}

This radial velocity for the runaway star supports the notion that HIP 47155 is a former companion candidate to \psrjns{} (\autoref{fig:0630HIP}). The predicted NS properties and time and place of the supernova event are given in \autoref{tab:0630_47155}. The position is close to the centre of the Antlia SNR. We therefore conclude that \psrjns{} and HIP 47155 were ejected in the same supernova that formed the Antlia SNR $\approx\unit[1{.}2]{Myr}$ ago at a distance to Earth of $\approx\unit[120]{pc}$.

\begin{table}
\centering
\caption[Predicted current parameters of \psrjns{} and supernova position and time.]{Predicted current parameters of \psrjns{} and supernova position and time for those runs where \psrjns{} and HIP 47155 were within the Antlia SNR.\newline
Predicted NS parameters: heliocentric radial velocity $v_r$, parallax $\pi$, proper motion $\mu_{\alpha}^*$, $\mu_\delta$, peculiar space velocity $v_{sp}$; Predicted supernova position: distance of the supernova to Earth at the time of the supernova ($\dsunSN$) and as seen today ($\dsunheute$), Galactic coordinates (Galactic longitude $l$, Galactic latitude $b$, J2000.0) as seen from the Earth today; Predicted time of the supernova in the past $\tau$. For the deduction of the parameters we refer to \citealt{2010MNRAS.402.2369T}.}\label{tab:0630_47155}
{
\begin{tabular}{l  c}
\toprule
\multicolumn{2}{p{0.35\textwidth}<{\centering}}{Predicted present-day parameters of \psrjns{}}\\\midrule
$v_r$ [km/s]             	& $177^{+79}_{-20}$	\\
$\pi$ [mas]       	& $2{.}9^{+0{.}4}_{-0{.}3}$\\
$v_{sp}$ [km/s]					 		& $177^{+76}_{-21}$			\\\midrule
\multicolumn{2}{p{0.35\textwidth}<{\centering}}{Predicted supernova/SNR position and time}\\\midrule
$\dsunSN$ [pc] & $128^{+17}_{-17}$\\
$\dsunheute$ [pc]  & $138^{+17}_{-17}$ \\
$l$ [$\mathrm{^\circ}$]     & $270{.}4^{+1{.}9}_{-1{.}9}$	\\
$b$ [$\mathrm{^\circ}$]     & $19{.}2^{+0{.}5}_{-3{.}0}$	\\
$\tau$ [Myr]   	& $1{.}20^{+0{.}26}_{-0{.}18}$	\\\bottomrule
\end{tabular}
}
\end{table}
%


\section{Conclusions}\label{sec:concl}

We searched for the NS that was born in the same supernova event that formed the Antlia SNR. In order to account for the uncertainties in the observables as well as the unknown radial velocity of NSs, we performed Monte Carlo simulations and evaluated the outcome statistically. 

\begin{figure}
\centering
\fbox{\includegraphics[clip, width=0.3\textwidth, viewport= 123 267 470 592]{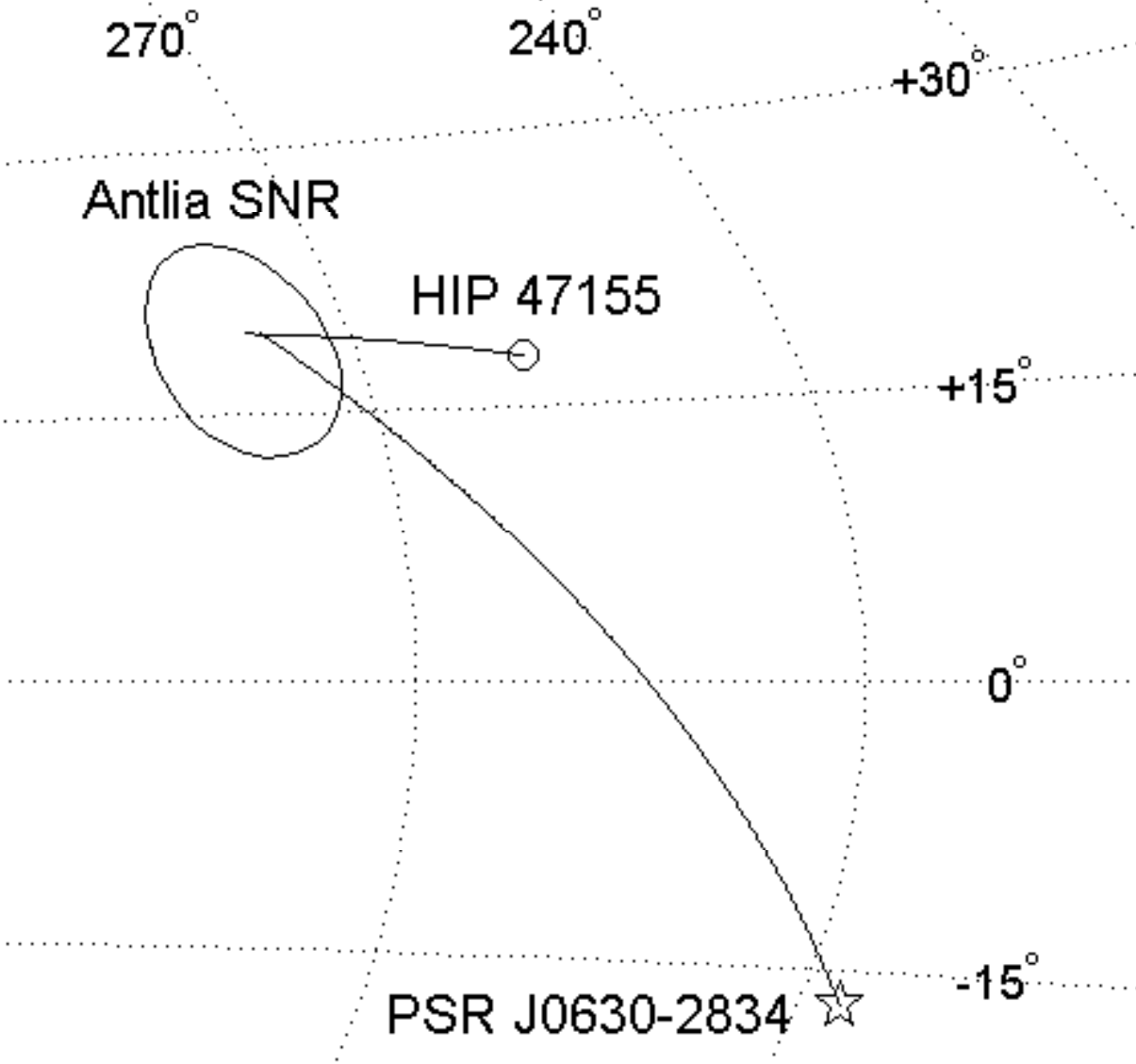}}
\caption{Past trajectories for \psrjns{} and HIP 47155 projected on a Galactic coordinate system (for a particular set of input parameters). Present positions are marked with a star for the neutron star and a circle for the runaway star. The large circle denotes the Antlia SNR with its present extension (but at the position at the time of the supernova).}
\label{fig:pastflightpaths}
\end{figure}

The most promising candidate for an association with the Antlia SNR is the young pulsar \psrjns{} (\autoref{fig:pastflightpaths}). 
The predicted kinematic age of the NS, i.e. age of the SNR is $\approx\unit[1{.}2]{Myr}$. This kinematic age is in reasonable agreement with the pulsar's spin-down age of $\unit[2{.}77]{Myr}$ \citep{2004MNRAS.353.1311H}. Discrepancies between the spin-down and kinematic pulsar ages of a factor of a few are well known, in particular for young and middle-aged pulsars \citep[e.g.][]{2001ApJ...554..161P,2002ApJ...567L.141M,2011MNRAS.417..617T}.

The runaway star HIP 47155 is suggested to be a former companion to \psrjns{}, and has been classified as a $\lambda$ Bootes star \citep[kA3hF2mA3V,][]{2001A&A...373..633P}. We have determined its radial velocity to be $\unit[32{.}42\pm0{.}70]{km/s}$. Although the majority of the $\lambda$ Bootes stars are believed to be old ($\approx$Gyr), a few per cent of them are considerably younger (a few tens of Myr, \citealt{2002MNRAS.336.1030P}), some having been found in the very young Orion OB1 association and the cluster NGC 2264 \citep{2001A&A...373..633P}. Thus, the young age we infer for HIP 47155 is not inconsistent with its nature and it remains a possible former companion to \psrjns{}.

In this scenario, the distance of the Antlia SNR is $\approx\unit[135]{pc}$. This distance is also in agreement with that the Antlia SNR lies in the region of the Vela SNR and the Gum nebula or just in front of it. The Antlia SNR adjoins the Gum nebula \citep{2007ApJ...670.1132S}.

Integrating the COMPTEL $\unit[1{.}8]{MeV}$ flux \citep{2001ESASP.459...55P,2010A&A...522A..51D} over a circle centred at $\left(l,b\right)={\left(270\commacirc4^{+1\commacirc9}_{-1\commacirc9},19\commacirc2^{+0\commacirc5}_{-3\commacirc0}\right)}$ with a radius of $\unit[12]{^\circ}$ and adopting a distance of $\unit[{138^{+17}_{-17}}]{pc}$ yields a mass of $^{26}$Al that was ejected during the supernova of $\approx\unit[{5{.}2\pm1{.}6\cdot10^{-5}}]{M_\odot}$. Compared to theoretical $^{26}$Al yields (from core-collapse supernovae) by \citet{1995ApJS..101..181W} and \citet{2005NuPhA.758...11L}, this corresponds to a mass of the progenitor star of $\approx\unit[{14-32}]{M_\odot}$ (earlier than B1 on the main sequence, \citealt{Schmidt-Kaler1982,2010AN....331..349H}). 

The former host OB association or stellar cluster of the supernova progenitor probably lies in the Vela region. The massive progenitor possibly got ejected from its birth place prior its supernova due to dynamical interactions \citep[see e.g.][]{1989AJ.....98..217L}. It is known that about $20$ to $30$ per cent of supernova progenitors are located outside stellar groups \citep[e.g.][]{1998AJ....115..821M,2004ApJS..151..103M}.

At a supernova distance of $\approx\unit[110-150]{pc}$ about $\unit[1{.}0-1{.}5]{Myr}$ ago, this supernova may not be the only candidate for the supernova that produced the $^{60}$Fe found on Earth \citep{2004PhRvL..93q1103K,2008PhRvL.101l1101F}, as it might be too distant and/or too recent. However, this supernova may have produced some of the $^{60}$Fe found and contributed to reheating the Local Bubble.

\section*{Acknowledgments}

We are grateful to P.\ Berlind and G.\ Esquerdo for help with the
spectroscopic observations of HIP\,47155. We thank R.\ Diehl for kindly providing us the COMPTEL $\unit[1{.}8]{MeV}$ map and fruitful discussions. We would also like to thank the anonymous referee for useful comments.

NT acknowledges Carl-Zeiss-Stiftung for a scholarship.
GT acknowledges partial
support for this work from NSF grant AST-1007992.
NT, RN and MMH acknowledge support from DFG in the SFB/TR-7 Gravitational
Wave Astronomy.
This work has made use of the Australia Telescope National Facility (ATNF) pulsar database and the Simbad, operated
at the Centre de Donn\'ees astronomiques de Strasbourg (CDS).
\bibliographystyle{mn2e}
\bibliography{bib_Fe60Paper}

\begin{thebibliography}{}

\bibitem[\protect\citeauthoryear{{Blaauw}}{{Blaauw}}{1961}]{1961BAN....15..265B}
{Blaauw} A.,  1961, \bain, 15, 265

\bibitem[\protect\citeauthoryear{{Blinnikov}, {Imshennik} \&
  {Utrobin}}{{Blinnikov} et~al.}{1982}]{1982SvAL....8..361B}
{Blinnikov} S.~I.,  {Imshennik} V.~S.,    {Utrobin} V.~P.,  1982, Soviet
  Astronomy Letters, 8, 361

\bibitem[\protect\citeauthoryear{{Bobylev}}{{Bobylev}}{2008}]{2008AstL...34..686B}
{Bobylev} V.~V.,  2008, Astronomy Letters, 34, 686

\bibitem[\protect\citeauthoryear{{Bobylev} \& {Bajkova}}{{Bobylev} \&
  {Bajkova}}{2009}]{2009AstL...35..396B}
{Bobylev} V.~V.,  {Bajkova} A.~T.,  2009, Astronomy Letters, 35, 396

\bibitem[\protect\citeauthoryear{{Brisken}, {Benson}, {Goss} \&
  {Thorsett}}{{Brisken} et~al.}{2002}]{2002ApJ...571..906B}
{Brisken} W.~F.,  {Benson} J.~M.,  {Goss} W.~M.,    {Thorsett} S.~E.,  2002,
  \apj, 571, 906

\bibitem[\protect\citeauthoryear{{Buchhave}}{{Buchhave}}{2010}]{Buchhave:10a}
{Buchhave} L.~A.,  2010, PhD thesis, University of Copenhagen, Denmark

\bibitem[\protect\citeauthoryear{{Buchhave}, {Bakos}, {Hartman}, {Torres},
  {Kov{\'a}cs}, {Latham}, {Noyes}, {Esquerdo}, {Everett}, {Howard} \&
  {et~al.}}{{Buchhave} et~al.}{2010}]{Buchhave:10b}
{Buchhave} L.~A.,  {Bakos} G.~{\'A}.,  {Hartman} J.~D.,  {Torres} G.,
  {Kov{\'a}cs} G.,  {Latham} D.~W.,  {Noyes} R.~W.,  {Esquerdo} G.~A.,
  {Everett} M.,  {Howard}   {et~al.} 2010, \apj, 720, 1118

\bibitem[\protect\citeauthoryear{{Buchhave}, {Latham}, {Johansen}, {Bizzarro},
  {Torres}, {Rowe}, {Batalha}, {Borucki}, {Brugamyer}, {Caldwell} \&
  {et~al.}}{{Buchhave} et~al.}{2012}]{2012Natur.486..375B}
{Buchhave} L.~A.,  {Latham} D.~W.,  {Johansen} A.,  {Bizzarro} M.,  {Torres}
  G.,  {Rowe} J.~F.,  {Batalha} N.~M.,  {Borucki} W.~J.,  {Brugamyer} E.,
  {Caldwell} C.,    {et~al.} 2012, \nat, 486, 375

\bibitem[\protect\citeauthoryear{{Casagrande}, {Sch{\"o}nrich}, {Asplund},
  {Cassisi}, {Ram{\'{\i}}rez}, {Mel{\'e}ndez}, {Bensby} \&
  {Feltzing}}{{Casagrande} et~al.}{2011}]{2011A&A...530A.138C}
{Casagrande} L.,  {Sch{\"o}nrich} R.,  {Asplund} M.,  {Cassisi} S.,
  {Ram{\'{\i}}rez} I.,  {Mel{\'e}ndez} J.,  {Bensby} T.,    {Feltzing} S.,
  2011, \aap, 530, A138

\bibitem[\protect\citeauthoryear{{Deller}, {Tingay}, {Bailes} \&
  {Reynolds}}{{Deller} et~al.}{2009}]{2009ApJ...701.1243D}
{Deller} A.~T.,  {Tingay} S.~J.,  {Bailes} M.,    {Reynolds} J.~E.,  2009,
  \apj, 701, 1243

\bibitem[\protect\citeauthoryear{{Diehl}, {Lang}, {Martin}, {Ohlendorf},
  {Preibisch}, {Voss}, {Jean}, {Roques}, {von Ballmoos} \& {Wang}}{{Diehl}
  et~al.}{2010}]{2010A&A...522A..51D}
{Diehl} R.,  {Lang} M.~G.,  {Martin} P.,  {Ohlendorf} H.,  {Preibisch} T.,
  {Voss} R.,  {Jean} P.,  {Roques} J.,  {von Ballmoos} P.,    {Wang} W.,  2010,
  \aap, 522, A51+

\bibitem[\protect\citeauthoryear{{Eisenbeiss}}{{Eisenbeiss}}{2011}]{ThomasPhD}
{Eisenbeiss} T.,  2011, PhD thesis, submitted, AIU,
  Friedrich-Schiller-Universit\"at Jena, Germany

\bibitem[\protect\citeauthoryear{{Fern{\'a}ndez}, {Figueras} \&
  {Torra}}{{Fern{\'a}ndez} et~al.}{2008}]{2008A&A...480..735F}
{Fern{\'a}ndez} D.,  {Figueras} F.,    {Torra} J.,  2008, \aap, 480, 735

\bibitem[\protect\citeauthoryear{{F\H{u}r\'esz}}{{F\H{u}r\'esz}}{2008}]{Furesz08}
{F\H{u}r\'esz} G.,  2008, PhD thesis, University of Szeged, Hungary

\bibitem[\protect\citeauthoryear{{Fitoussi}, {Raisbeck}, {Knie} \& {et
  al.}}{{Fitoussi} et~al.}{2008}]{2008PhRvL.101l1101F}
{Fitoussi} C.,  {Raisbeck} G.~M.,  {Knie} K.,    {et al.} 2008, Physical Review
  Letters, 101, 121101

\bibitem[\protect\citeauthoryear{{Hobbs}, {Lorimer}, {Lyne} \&
  {Kramer}}{{Hobbs} et~al.}{2005}]{2005MNRAS.360..974H}
{Hobbs} G.,  {Lorimer} D.~R.,  {Lyne} A.~G.,    {Kramer} M.,  2005, \mnras,
  360, 974

\bibitem[\protect\citeauthoryear{{Hobbs}, {Lyne}, {Kramer}, {Martin} \&
  {Jordan}}{{Hobbs} et~al.}{2004}]{2004MNRAS.353.1311H}
{Hobbs} G.,  {Lyne} A.~G.,  {Kramer} M.,  {Martin} C.~E.,    {Jordan} C.,
  2004, \mnras, 353, 1311

\bibitem[\protect\citeauthoryear{{Hohle}, {Neuh{\"a}user} \& {Schutz}}{{Hohle}
  et~al.}{2010}]{2010AN....331..349H}
{Hohle} M.~M.,  {Neuh{\"a}user} R.,    {Schutz} B.~F.,  2010, Astronomische
  Nachrichten, 331, 349

\bibitem[\protect\citeauthoryear{{Hoogerwerf}, {de Bruijne} \& {de
  Zeeuw}}{{Hoogerwerf} et~al.}{2001}]{2001A&A...365...49H}
{Hoogerwerf} R.,  {de Bruijne} J.~H.~J.,    {de Zeeuw} P.~T.,  2001, \aap, 365,
  49

\bibitem[\protect\citeauthoryear{{Houk}}{{Houk}}{1982}]{1982mcts.book.....H}
{Houk} N.,  1982, {Michigan Catalogue of Two-dimensional Spectral Types for the
  HD stars. Volume 3. Declinations $-40^\circ$ to $-26^\circ$.}

\bibitem[\protect\citeauthoryear{{Kaplan}, {van Kerkwijk} \&
  {Anderson}}{{Kaplan} et~al.}{2007}]{2007ApJ...660.1428K}
{Kaplan} D.~L.,  {van Kerkwijk} M.~H.,    {Anderson} J.,  2007, \apj, 660, 1428

\bibitem[\protect\citeauthoryear{{Knie}, {Korschinek}, {Faestermann}, {Dorfi},
  {Rugel} \& {Wallner}}{{Knie} et~al.}{2004}]{2004PhRvL..93q1103K}
{Knie} K.,  {Korschinek} G.,  {Faestermann} T.,  {Dorfi} E.~A.,  {Rugel} G.,
  {Wallner} A.,  2004, Physical Review Letters, 93, 171103

\bibitem[\protect\citeauthoryear{{Leonard}}{{Leonard}}{1989}]{1989AJ.....98..217L}
{Leonard} P.~J.~T.,  1989, \aj, 98, 217

\bibitem[\protect\citeauthoryear{{Limongi} \& {Chieffi}}{{Limongi} \&
  {Chieffi}}{2005}]{2005NuPhA.758...11L}
{Limongi} M.,  {Chieffi} A.,  2005, Nuclear Physics A, 758, 11

\bibitem[\protect\citeauthoryear{{Ma{\'{\i}}z-Apell{\'a}niz}, {Walborn},
  {Galu{\'e}} \& {Wei}}{{Ma{\'{\i}}z-Apell{\'a}niz}
  et~al.}{2004}]{2004ApJS..151..103M}
{Ma{\'{\i}}z-Apell{\'a}niz} J.,  {Walborn} N.~R.,  {Galu{\'e}} H.~A.,    {Wei}
  L.~H.,  2004, \apjs, 151, 103

\bibitem[\protect\citeauthoryear{{Manchester}, {Hobbs}, {Teoh} \&
  {Hobbs}}{{Manchester} et~al.}{2005}]{2005AJ....129.1993M}
{Manchester} R.~N.,  {Hobbs} G.~B.,  {Teoh} A.,    {Hobbs} M.,  2005, \aj, 129,
  1993

\bibitem[\protect\citeauthoryear{{Mason}, {Gies}, {Hartkopf}, {Bagnuolo} Jr.,
  {ten Brummelaar} \& {McAlister}}{{Mason} et~al.}{1998}]{1998AJ....115..821M}
{Mason} B.~D.,  {Gies} D.~R.,  {Hartkopf} W.~I.,  {Bagnuolo} Jr. W.~G.,  {ten
  Brummelaar} T.,    {McAlister} H.~A.,  1998, \aj, 115, 821

\bibitem[\protect\citeauthoryear{{McCullough}, {Fields} \&
  {Pavlidou}}{{McCullough} et~al.}{2002}]{2002ApJ...576L..41M}
{McCullough} P.~R.,  {Fields} B.~D.,    {Pavlidou} V.,  2002, \apjl, 576, L41

\bibitem[\protect\citeauthoryear{{Migliazzo}, {Gaensler}, {Backer}, {Stappers},
  {van der Swaluw} \& {Strom}}{{Migliazzo} et~al.}{2002}]{2002ApJ...567L.141M}
{Migliazzo} J.~M.,  {Gaensler} B.~M.,  {Backer} D.~C.,  {Stappers} B.~W.,  {van
  der Swaluw} E.,    {Strom} R.~G.,  2002, \apjl, 567, L141

\bibitem[\protect\citeauthoryear{{Motch}, {Zavlin} \& {Haberl}}{{Motch}
  et~al.}{2003}]{2003AA...408..323M}
{Motch} C.,  {Zavlin} V.~E.,    {Haberl} F.,  2003, \aap, 408, 323

\bibitem[\protect\citeauthoryear{{Paunzen}}{{Paunzen}}{2001}]{2001A&A...373..633P}
{Paunzen} E.,  2001, \aap, 373, 633

\bibitem[\protect\citeauthoryear{{Paunzen}, {Duffee}, {Heiter}, {Kuschnig} \&
  {Weiss}}{{Paunzen} et~al.}{2001}]{2001A&A...373..625P}
{Paunzen} E.,  {Duffee} B.,  {Heiter} U.,  {Kuschnig} R.,    {Weiss} W.~W.,
  2001, \aap, 373, 625

\bibitem[\protect\citeauthoryear{{Paunzen}, {Iliev}, {Kamp} \&
  {Barzova}}{{Paunzen} et~al.}{2002}]{2002MNRAS.336.1030P}
{Paunzen} E.,  {Iliev} I.~K.,  {Kamp} I.,    {Barzova} I.~S.,  2002, \mnras,
  336, 1030

\bibitem[\protect\citeauthoryear{{Pivovaroff}, {Kaspi}, {Camilo}, {Gaensler} \&
  {Crawford}}{{Pivovaroff} et~al.}{2001}]{2001ApJ...554..161P}
{Pivovaroff} M.~J.,  {Kaspi} V.~M.,  {Camilo} F.,  {Gaensler} B.~M.,
  {Crawford} F.,  2001, \apj, 554, 161

\bibitem[\protect\citeauthoryear{{Pl{\"u}schke}, {Diehl}, {Sch{\"o}nfelder},
  {Bloemen}, {Hermsen}, {Bennett}, {Winkler}, {McConnell}, {Ryan}, {Oberlack}
  \& {Kn{\"o}dlseder}}{{Pl{\"u}schke} et~al.}{2001}]{2001ESASP.459...55P}
{Pl{\"u}schke} S.,  {Diehl} R.,  {Sch{\"o}nfelder} V.,  {Bloemen} H.,
  {Hermsen} W.,  {Bennett} K.,  {Winkler} C.,  {McConnell} M.,  {Ryan} J.,
  {Oberlack} U.,    {Kn{\"o}dlseder} J.,  2001, in {Gimenez} A.,  {Reglero} V.,
    {Winkler} C.,  eds, Exploring the Gamma-Ray Universe Vol.~459 of ESA
  Special Publication, {The COMPTEL 1.809 MeV survey}.
pp 55--58

\bibitem[\protect\citeauthoryear{{Schmidt-Kaler}}{{Schmidt-Kaler}}{1982}]{Schmidt-Kaler1982}
{Schmidt-Kaler} T.~H.,  1982, {Physical parameters of the stars}

\bibitem[\protect\citeauthoryear{{Shinn}, {Min}, {Sankrit}, {Ryu}, {Kim},
  {Han}, {Nam}, {Park}, {Edelstein} \& {Korpela}}{{Shinn}
  et~al.}{2007}]{2007ApJ...670.1132S}
{Shinn} J.-H.,  {Min} K.~W.,  {Sankrit} R.,  {Ryu} K.-S.,  {Kim} I.-J.,  {Han}
  W.,  {Nam} U.-W.,  {Park} J.-H.,  {Edelstein} J.,    {Korpela} E.~J.,  2007,
  \apj, 670, 1132

\bibitem[\protect\citeauthoryear{{Tetzlaff}}{{Tetzlaff}}{2013}]{2013PhD...Nina}
{Tetzlaff} N.,  2013, PhD thesis, AIU, Friedrich-Schiller-Universit\"at Jena,
  Germany

\bibitem[\protect\citeauthoryear{{Tetzlaff}, {Eisenbeiss}, {Neuh{\"a}user} \&
  {Hohle}}{{Tetzlaff} et~al.}{2011}]{2011MNRAS.417..617T}
{Tetzlaff} N.,  {Eisenbeiss} T.,  {Neuh{\"a}user} R.,    {Hohle} M.~M.,  2011,
  \mnras, 417, 617

\bibitem[\protect\citeauthoryear{{Tetzlaff}, {Neuh{\"a}user} \&
  {Hohle}}{{Tetzlaff} et~al.}{2009}]{2009MNRAS.400L..99T}
{Tetzlaff} N.,  {Neuh{\"a}user} R.,    {Hohle} M.~M.,  2009, \mnras, 400, L99

\bibitem[\protect\citeauthoryear{{Tetzlaff}, {Neuh{\"a}user} \&
  {Hohle}}{{Tetzlaff} et~al.}{2011}]{2011MNRAS.410..190T}
{Tetzlaff} N.,  {Neuh{\"a}user} R.,    {Hohle} M.~M.,  2011, \mnras, 410, 190

\bibitem[\protect\citeauthoryear{{Tetzlaff}, {Neuh{\"a}user}, {Hohle} \&
  {Maciejewski}}{{Tetzlaff} et~al.}{2010}]{2010MNRAS.402.2369T}
{Tetzlaff} N.,  {Neuh{\"a}user} R.,  {Hohle} M.~M.,    {Maciejewski} G.,  2010,
  \mnras, 402, 2369

\bibitem[\protect\citeauthoryear{{Tetzlaff}, {Schmidt}, {Hohle} \&
  {Neuh{\"a}user}}{{Tetzlaff} et~al.}{2012}]{2012PASA...29...98T}
{Tetzlaff} N.,  {Schmidt} J.~G.,  {Hohle} M.~M.,    {Neuh{\"a}user} R.,  2012,
  \pasa, 29, 98

\bibitem[\protect\citeauthoryear{{Torres}, {Quast}, {Melo} \&
  {Sterzik}}{{Torres} et~al.}{2008}]{2008hsf2.book..757T}
{Torres} C.~A.~O.,  {Quast} G.~R.,  {Melo} C.~H.~F.,    {Sterzik} M.~F.,  2008,
  {Young Nearby Loose Associations}.
pp 757--+

\bibitem[\protect\citeauthoryear{{Ugliano}, {Janka}, {Marek} \&
  {Arcones}}{{Ugliano} et~al.}{2012}]{2012ApJ...757...69U}
{Ugliano} M.,  {Janka} H.-T.,  {Marek} A.,    {Arcones} A.,  2012, \apj, 757,
  69

\bibitem[\protect\citeauthoryear{{Woosley} \& {Weaver}}{{Woosley} \&
  {Weaver}}{1995}]{1995ApJS..101..181W}
{Woosley} S.~E.,  {Weaver} T.~A.,  1995, \apjs, 101, 181

\end{thebibliography}

\label{lastpage}

\end{document}